\documentclass[12pt,twocolumn]{article}
\usepackage{graphicx}
\usepackage{color}
\usepackage{hyperref}
\usepackage{caption}
\usepackage{setspace}
\usepackage[margin=1in]{geometry}
\usepackage{ulem}
\usepackage[T1]{fontenc}
\usepackage{amsmath}
\usepackage[utf8]{inputenc}
\usepackage{hyperref}
\usepackage{epstopdf}
\onecolumn
\begin{document}
\begin{center}
{\Large {\bf Critical Phenomena Study of 3D Heisenberg  Magnet}}
\end{center}
\vskip 1cm
\begin{center}
{\it Nepal Banerjee}\\
{\it Department of Physics,University of Seoul,South korea}
\vskip 1 cm
{\it Email:\ nb.uos1989@gmail.com}\\
\end{center}
\vskip 1 cm
\begin{abstract}

Recent discovery of several van der waals magnetic material and moire magnet introduce to us an extremely challenging and revolutionary era of 2D magnetism and correlated phenomena for low dimensional material\cite{ajayan2016van}\cite{jung2015origin}\cite{song}\cite{blanet}\cite{chen2019signatures}\cite{chen2019evidence}\cite{tran2019evidence}\cite{jung2014ab}\cite{tran2019evidence}\cite{xie2023evidence}\cite{wu2017topological}\cite{andrei2021marvels}\cite{burch2018magnetism}\cite{gong2017discovery}\cite{park2016opportunities}\cite{jiang2021recent}.
More often the simplest spin models which is based on inter-atomic exchange and spin-orbit coupling(SOC) potentially able to capture and  explain the critical phenomena of extremely complicated correlated magnetic material\cite{anderson}\cite{anderson2}\cite{PhysRevLett.62.1694}\cite{sachdev1999quantum}\cite{lado}\cite{rehn2016classical}\cite{auerbach}\cite{khomskii}.In this work we have attempted to simulate 3D Heisenberg magnet using classical Monte Carlo simulation\cite{binder,sandvik,chen,mallick}.Our goal is to establish a new and simplest spin simulation technique which can help us to understand those van der waals magnet from its microscopic length scale.Here we have been proposing a completely new methodology of classical Monte Carlo simulation of Heisenberg spin which is based on single spin flipping  Metropolis algorithm\cite{acharyya2005}.Our state of art simulation technique potentially able to study the phase transition of isotropic  XY(O(2)) and XYZ(O(3))spin model very efficiently.With this simulation technique we overcome the barrier of critical slowing down during the phase transition in a effective way and able to predict the transition temperature($T_c$) very accurately\cite{bishop}\cite{MA}\cite{Hasen}\cite{bishop2}\cite{kosterlitz1}\cite{kosterlitz2}\cite{peter}.
\end{abstract}
\vskip 1 cm
\twocolumn
\section{Introduction}
In theoretical condensed matter physics and statistical physics isotropic XY spin model and isotropic XYZ spin model is one of the fundamental model and it has been studied in a extensive way\cite{Fisher}\cite{cardy}\cite{Kardar}\cite{suzuki}\cite{brown1996high}\cite{holm1993finite}\cite{holm1993critical}\cite{holm1997critical} \cite{peczak1991high}\cite{Bikas}\cite{Rajesh}
\cite{rajesh2}.The lattice version of these models,which is known and popular as Bosonic Hubbard model has been widely used as the basic model for studying the strongly correlated phase of matter.Also we know that the phase transition of XY spin model and Bosonic fluid follows the same universality class.More compactly we can say that  this simple but fundamental models work as a common play ground for several correlated phase of matter,like quantum magnet,skyrmion,chiral magnets,superconductor,optical lattice for last several decades\cite{chalker1}\cite{chalker2}\cite{chalker3}\cite{dasgupta}\cite{sachdev2018}\cite{jalabert1991spontaneous}\cite{xu2008ising} \cite{sachdev1999translational}\cite{samajdar}\cite{dsen1}\cite{dsen2}\cite{mol}\cite{randeria}\cite{sujay}\cite{tomadin}.Last few years the study of 2D and 3D Heisenberg magnet becomes more relevant and an important field of research in the contest of van der waals material,skyrmionic phase,spintronics,topological phase of van der waals magnet\cite{nov1}\cite{nov2}\cite{nov3}\cite{nov4}\cite{nepal_topo_2}
\cite{sinova2015spin}\cite{sinova2004universal}\cite{nagaosa2010anomalous}\cite{jungwirth2006theory}\cite{gao2021layer}\cite{bernevig2022progress}\cite{kubler2014non}\cite{jena2020elliptical}\cite{vojta2005quantum}.This van der waals magnet has a wonderful properties and because of the invention of several sophisticated experimental technique this kind of van der waals magnet is easily achievable now a days in laboratory.The discovery of graphene has  prompted and show the pathway to invent different types of 2D and layer van der waals magnetic material\cite{nov5}\cite{nov6}\cite{nov9}.This van der waals materials are usually achievable with doping metal chalcogen atom inside a semiconducting material and it gives us a real scenario and rich platform to study the properties of those interesting material.It also become a great challenge for both theoretically and numerically to explain the different complex properties of those material using  simple theoretical model and simulation.Recently the discovery of several  TMD material and their twisted bilayer version with formula $MX_2$,where M=transition metal and X= chalcogen atom,transition metal trihalide with formula $MX_3$,where M=transition metal and X=halide and transition metal trichalcogenide with formula $MAX_3$,M=transition metal,
X=chalcogen and A=semiconducting material are  a big achievement of this field and this all materials act as an alternate material of graphene and hosting plethora of interesting correlated phase\cite{chit1}\cite{chit2}\cite{PhysRevLett.121.266401}\cite{PhysRevB.102.075413}\cite{PhysRevB.103.L121102}\cite{manzeli20172d}\cite{devakul2021magic}\cite{beniwal2023influence}.Specially monolayer and bilayer $CrI_3$ act as a promising platform to simulate the physics which arises because of strong magneto-electric effect in that material\cite{cri3_1}\cite{cri3_2}\cite{cri3_3}\cite{cri3_4}
\cite{cri3_5}\cite{cri3_6}.$MoTe_2$ and $WTe_2$ shows type-II weyl semimetalic phase and several correlated phase of matter emerges from bilayer and twisted bilayer version of this materials.This type of 2D TMD material open up the door of infinite possibility of hosting plethora of interesting phase.This van der waals material act as real platform and perfect test ground for well established theoretical spin model and that only become possible because of its high tunability\cite{Onsager}\cite{PA_Lee}\cite{kitaev}\cite{Elmers}.According to Mermin and Wagner theorem spontaneous symmetry breaking grounds state is restricted in a two dimensional isotropic Heisenberg model\cite{Mermin_wagner}.This theorem act as basic guiding principle for several decades in the direction of advancement of the  quantum field theoretical model of those two dimensional magnetic system.But often this theorem is misunderstood thinking that any order state is excluded in two dimension by this theorem.But this theorem strictly says that in presence of isotropic environment and presence of short-range interaction we can't expect any order state from two dimensional isotropic Heisenberg spin model at any finite T.Thermal fluctuation of isotropic Heisenberg spin destroy the order state at any finite T.During that time Peierls showed that how the localization of particle is destroyed by long-wavelength lattice wave in his seminal work and after that Bloch also proved using spin wave theory that the spontaneous magnetization is destroyed by long-wavelength spin waves.Both of them show that deviation from the order state is proportional to $\int \frac{d^d k}{k^2}$.Using straight forward calculation we can show that in 3D this integration converge to a non-zero finite value.But in 2D case for lower limit (1/L) this deviation turn into ln(L) and naturally diverge with L\cite{tobochnik1979monte}.The noble and breakthrough work done by Kosterlitz and Thouless first shows using XY spin model that how order state could developed in two dimension using continuous type of Heisenberg spin model.Kosterlitz and Thouless in their pioneer work shows that how the topological defect,which is here vortex and anti-vortex state can trigger the spontaneous symmetry breaking of ground state in continuous 2D XY spin model system\cite{cao}\cite{arnold}\cite{Berezinsky:1970fr}\cite{kosterlitz1973ordering}\cite{stryjewski1977metamagnetism}\cite{acharyya2015ising}\cite{kosterlitz1974critical}\cite{krech1999spin}.
The intrinsic anisotropic field is the major obstacle for simulating the isotropic environment for a Heisenberg spin system and because of that it is really a big challenge for both numerically and experimentally to simulate an isotropic environment.More often in material because of crystal asymmetry and present of intrinsic magnetic moment the Heisenberg spin prefer a biased direction for settled its ground state and with out any external field we observed a spontaneous symmetry breaking.Also another important factor which is hindering to simulate the pure isotropic environment is the finite size of the real lattice system.Because of that finite size effect we always have a residual magnetic moment which is more often grow up as an anisotropic field during the phase transition and end up with triggering  the spontaneous symmetry breaking along a particular direction and finally spin system choose an order ground state to settle down at low temperature.The seminal work done by Watson,Blume and Vineyard also observed order ground state with 2D isotropic Heisenberg model and later this work is extended by Brown and Ciftan also observed same kind of results and after that  Lau and Dasgupta pointed out that the topological point defect is the main reason for appearing order state in 2D and 3D isotropic Heisenberg spin system at low temperature.Brown and Ciftan showed that to get vanishing order state at finite T we need lattice size $L \sim 10^{300}$ for 2D Heisenberg spin.In their study they founded Tc=0.6 for $150 \times 150$ lattice which they classified as KT transition \cite{brown19932d}\cite{WBV}\cite{WBV2}\cite{chandan}\cite{Elton}\cite{ritchie1972theory}.\ Recently several study has reported about this consequences which usually appears because of the finite size effect of the lattice system.We have also founded order state in van der waals material in presence of short-range interaction in 2D.But at very low temperature  single spin flipping Metropolis technique doesn’t work properly and most of the time we end up with estimate wrong transition temperature(Tc).Here in this work we have mainly focus on 3D Heisenberg spin system which give us quite accurate results using classical Monte Carlo simulation with Heisenberg spin.So in this present work we have explored the basic feature of 3D XY and XYZ model using our simulation technique and study the critical behaviour of the spin system during the phase transition.Monte Carlo simulation with Heisenberg spin model is mostly performed by using cluster spin flipping algorithm and Swendsen-Wang and Wolff cluster flipping algorithm are most common algorithm for Monte Carlo simulation.But the implementation of cluster flipping algorithm is not only complicated but also difficult when we simulate complicated lattice system using that technique.Also the numerical haphazardness makes whole study not only very complex but also less intuitive for track down the whole simulation process.Here for simplicity we have been adopting single spin flipping Metropolis algorithm to equilibrate the random spin configuration.We use basic Monte Carlo step to simulate the numbers of spin configurations,which can acts as ensembles for statistical average of different thermodynamic quantities.The random initial spin is simulated after choosing randomly $\theta$ and $\phi$ coordinates in spherical polar co-ordinate system and we have simulated the spin components on unit sphere.The radius of that unit sphere indicate the magnitude of that classical spin moment.We have been observing very clearly phase transition in our present model using our state art single spin flipping Monte Carlo simulation technique.Also we have studied the critical phenomena using that proposed simulation technique.Using our simulation technique we able to study the variation of magnetization,susceptibility,energy and heat-capacity like macroscopic thermodynamic quantities with temperature.We have been observing a sharp peak on susceptibility and heat-capacity behaviour during this transition around $T_c=2.20$ for 3D XY and $T_c=1.449$ for 3D XYZ model and that peak grow up with system size.This behaviour generically indicates a second order phase transition in this present model and simulation.We have confirmed our observation after studying the scaling law.In this simulation we have been using the $5\times5\times5$ and $10\times10\times10$ cubic lattice grid to simulate the whole dynamics.Through out the classical Monte Carlo simulation we have used closed periodic boundary condition.We have planned our paper in the following way and presented our simulation results accordingly.We have presented and described the model Hamiltonian in section-2.In section-3 we have described the methodology and results of our simulation.In last section we have summarize the whole study as discussion and conclusion.    
\section{Model Hamiltonian}
\subsection{3D Isotropic XY Model}
Here we have considered isotropic ferromagnetic Heisenberg spin model.
We have considered XY type spin model system for our study.Here we  have represent our model Hamiltonian as 
\begin{eqnarray}
H =-J_x\sum_{<i,j>}S_x^i S_x^j -J_y\sum_{<i,j>}S_y^i S_y^j
\end{eqnarray}
We consider $J_x,J_y$ as a ferromagnetic coupling constant and $Jx >0,J_y>0 $.Here we consider isotropic version of this Hamiltonian so here $J_x=J_y=J$. Spin $S_i=(S_x,S_y)$ has $O(2)$ symmetry and has unit magnitude.We use $L\times L\times L$ simple cubic lattice grid for simulate the Heisenberg spin.Here we have included only nearest neighbour(NN) interaction and $<i,j>$ indicate the nearest neighbour interaction(NN) in a cubic lattice grid.We have considered $S_x=|S|cos(\phi)$ and $S_y=|S|s in(\phi)$ and for simplicity we consider J=1 and consider the magnitude of spin as |S|=1.Here each component of spin has three coordinate [i,j,k].We have simulated those spins in $L\times L \times L$  simple cubic lattice grid.Here we have considered classical spin and it is more likely act as compass,which are fixed at its own coordinate position and rotated in two dimensional spin space $[S_x,S_y]$.Here we have considered isotropic model and spins are equally prefer to orient itself along the both $S_x$ and $S_y$ direction.We have checked its anisotropic version where $J_x \neq J_y$.In that case spin will prefer to align  along a specific direction according to the strength of the exchange interaction.As an example we can say if the $J_x > J_y$ then spin will broke its ground  state symmetry along the $S_x$ direction since  $J_x$ is larger than $J_y$.We have studied the spin dynamics during the phase transition with T.Here for simplicity we set J=1,S=1 and consider Boltzman constant($k_B$) as 1.In this model we have observed a clear phase transition and  ground state is ferromagnetic.Here for 3D simple cubic lattice each spin has six nearest neighbour and they all interact with ferromagnetic exchange interaction.This model more often call as rotor model and its quantum version is studied in a extensive way.  

\subsection{3D Isotropic XYZ  Model}
Here we have considered the XYZ isotropic model for simulation.Here we have considered isotropic ferromagnetic interaction where J >0.We have considered spin components as $S_z=|S|cos(\theta)$,$S_x=|S|sin(\theta)cos(\phi)$, $S_y=|S|sin(\theta)sin(\phi)$.
\begin{equation}
\begin{split}
H=-J_x\sum_{<i,j>}S_x^i S_x^j &-J_y\sum_{<i,j>}S_y^i S_y^j\\
&-J_z\sum_{<i,j>}S_z^i S_z^j
\end{split}
\end{equation}
This spins are simulated on three dimensional simple cubic lattice so we can write spin as $S_x=S_x(i,j,k)$, $S_y=S_y(i,j,k)$,$S_z=S_z(i,j,k)$.Here <i,j> indicate nearest neighbour (NN) ferromagnetic interaction in simple cubic lattice grid.Here we consider isotropic Hamiltonian so $J_x=J_y=J_z=J$.Here we have considered J=1 and spin magnitude is |S|=1.Here spin has O(3) symmetry and its magnitude is one.Since here the interaction is isotropic so the spin are equally like to align along the $S_x$,$S_y$ and $S_z$ direction.Here for simplicity we consider J=1 and since J is positive so system will show the ferromagnetic ground state.Here we have simulated  spin in a simple cubic $L \times L \times L $ grid.So each spin has its three independent coordinate [i,j,k] in that spatial dimension.For simulating the random spin component we have chosen random $\theta$ and $\phi$ coordinate in spherical polar coordinate.Here we have considered the formula for that random $\theta$ and $\phi$ coordinate as $\theta=cos^{-1}(2u_1 -1)-\pi$ and $\phi=2\pi(1-u_2)$.Here $u_1$ and $u_2$ are computer generated random number and we tested its uniformity in advanced.

\section{Simulation Methodology and Results}
\subsection{3D XY Spin Model}
Here we will going to describe the methodology and results of our simulation.Here we have focused ourself strictly on isotropic XY Heisenberg spin model and this spin are simulated in 3D cubic lattice grid after choosing an angle $\phi$ randomly such a way that the random angle $\phi$ must be uniformly distributed between $0$ to $2\pi$.We have chosen random $\phi$ as $\phi=2\pi u_1$.Here $u_1$ is a computer generated random number and which is a tested as uniform random number generator.We have assigned the spin components of that XY spin as $S_x(i,j,k)=|S|cos(\phi)$ and $S_y(i,j,k)=|S|sin(\phi)$ in 3D cubic lattice grid,where $S_x,S_y$ are two component of that spin.The way we have chosen this components such that the  spin magnitude must be normalized to the value of |S|.Each spin has its independent coordinate (i,j,k) in that 3D cubic lattice grid.In that way we have simulated random spin configuration where all the spins has randomly chosen its $S_x$ and$S_y$  component and which are sitting on independent 3D cubic grid (i,j,k).We have assigned this random configuration of spins as a high temperature spin configuration,which is far above the transition temperature.Then we have started to cooling the system and decreasing the temperature with small step.In each step we have measured and calculate the different thermodynamic quantity like spontaneous magnetization,susceptibility,heat-capacity,energy.We have measured  all those thermodynamic quantity after taking ensembles average with $3\times10^4$ number of equivalent spin configuration at a particular temperature.Here we measure  spontaneous magnetization(M) which is square root average of magnetization of each component of a spin and we write it as a $M=\sqrt{(m_x^2 +m_y^2 )}$.Here $m_x$,$m_y$ are the average spin component along the $S_x,S_y$ direction respectively for each site of the lattice grid.At each temperature we take the average over $3\times10^4$ identical spin configuration which we call as ensembles for that particular temperature.We ignore $3\times 10^4$ ensembles just to reach equilibrium.Here we have calculated susceptibility,heat-capacity at each temperature using the same procedure mentioned before.We use the following formula for that calculation of susceptibility,which is $\chi=L^3(<M^2>-<M>^2)/T$ and for heat-capacity using the following formula $C_V=L^3(<E^2>-<E>^2)/T^2$. Here the energy is continuously decreasing with temperature and it is indicating that the system is going towards the reaching  of stable equilibrium state which is here ferromagnetic ground state.
\begin{figure}[htp]
\centering
\includegraphics[scale=0.20]{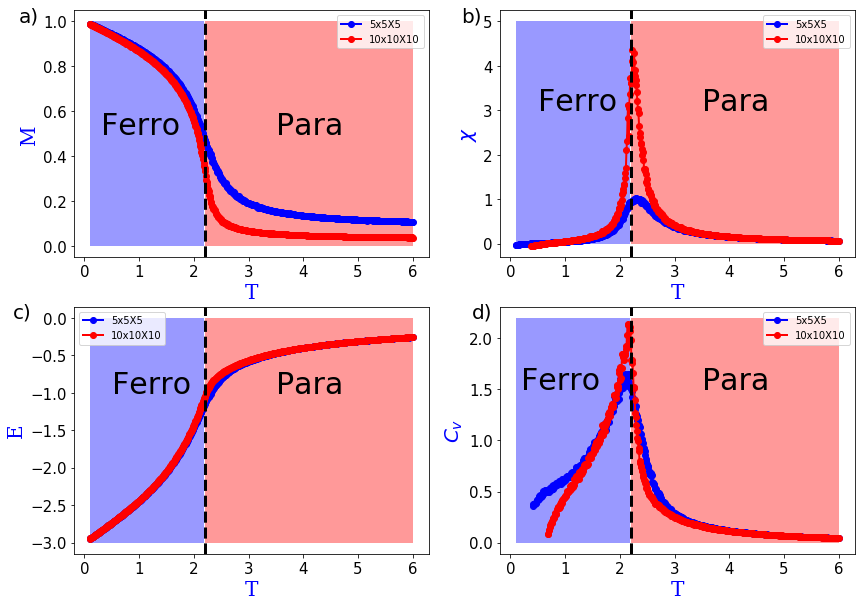}
\caption{a)Magnetization(M) with temperature (T). b)Susceptibility($\chi$) with temperature(T). c)Energy with temperature(T). d)Heat capacity($C_v$) with temperature(T).Here we presented this results for two different lattice size, which are $5 \times 5 \times 5 $, $10 \times 10 \times 10 $.Here we have presented the data for 3D XY Heisenberg spin model.Here dotted vertical line indicate the phase transition line.Here the transition temperature Tc=2.20.}
\label{}
\end{figure} 
Here we have estimated transition temperature ($T_c$)from susceptibility peak and heat-capacity peak and here that transition temperature is $T_c=2.20$,which is very closed to the other reported results in previous literature\cite{lou2007}\cite{canova2016}.Here we have studied the whole simulation for $5\times5\times5,10\times10\times10$ lattice site which is a cubic lattice grid.Here for a particular temperature we have equilibrate the spin configuration with $6\times 10^4$ MC steps among that we discard $3\times 10^4$ just to achieve equilibrium and consider $3\times 10^4$ steps for average after achieve that equilibrium spin configuration.In each step we use single spin flipping  Metropolis algorithm to equilibrate the spin configuration.This Metropolis flipping is the heart of the whole simulation process and we follow a very delicate but very simple and straight forward approach which is sometimes we call as continuous spin flipping methodology.So here we are going to discuss about that simulation methodology in details.Here we have selected randomly a site,let suppose it is (i,j,k) point.Then we have calculated the energy of that spin because of nearest neighbour interaction with its neighbour spin.Now we have created a new direction of spin state after choosing a random $\phi$ angle using the formula $\phi=2\pi(1- u_1) $.Here $u_1$ is a computer generated  random number,which we have already tested as a uniform random number generator.We have flipped the selected spin along that newly created random $(\phi )$ direction.Now we have calculated the energy of that flipped spin based on the interaction with its neighbour spin.Let suppose that before flipping the energy of that spin was E and after flipping the energy of that spin become $E_1$.Now we have  calculated the energy difference $\Delta=(E_1 -E)$ and using that difference we have calculated the flipping  probability using the following formula $P=e^{-\Delta/T}$.If that flipping probability P is greater then the random number then we have flipped that selected spin along the new direction otherwise we kept that spin in the previous state.We have selected another spin and repeat that same procedure and we have continue this selection process of spin site up to $L \times L \times L $ times so that each site get atleast a chance of being selected randomly.We call this method Metropolis flipping and we have done this process $6\times10^4$ times.Basically with this Metropolis flipping the spin configuration start to equilibrate slowly and create equilibrium unique spin configuration which we call as ensemble at a particular temperature and those equilibrium spin configuration is only considered for taking average of different thermodynamic variables.So finally we have set $3\times 10^4$ steps just to equilibrate the system using Metropolis flipping.Once the system reach at equilibrium then we have calculated different thermodynamic quantity using ensemble average.Here for taking average we use $3 \times 10^4$ ensembles.Here we have calculated the average value of $m_x$,$m_y$ with the formula $m_x=\sum S_x/L^3$ .Similarly we have calculated $m_y$.In this calculation we have calculated magnetization (M) using the formula $M=\sqrt{m_x^2 +m_y^2}$.We are observing that magnetization is changing from $10^{-2}$ oder to 1  continuously when temperature(T) is cool down from a higher temperature to lower temperature via the transition temperature $T=T_c$.We have observed  the final stable equilibrium spin configuration for which the  magnetization is settled down to 1 finally.
The variation of magnetization here sharply indicate a second order phase transition which goes through a continuous transition from the value of order $\sim 10^{-2}$ to 1.
Here we have studied two different lattice size which are $5\times 5 \times 5 $ and $10 \times 10 \times 10$ and we have observed finite size effect and verify the scaling law of our spin system.We also observe how the magnetization peak change with different system size.We have calculated average of $M^2$ which is denoted as $<M^2>$ and average of magnetization which we call $<M>$ at a particular T.This last two quantity is needed for calculate the susceptibility using the formula
\begin{eqnarray}
\chi=L^3(<M^2>-<M>^2)/T
\end{eqnarray} 
Similarly we have calculated the average value of energy of  each spin which we have denoted as $<E>$ and average value of $E^2$ which is $<E^2>$.After that we have calculated the heat-capacity using the formula
\begin{eqnarray}
C_V=L^3(<E^2>-<E>^2)/T^2
\end{eqnarray}   
Here the $\chi$ measure the fluctuation of magnetization which is diverge during the second order phase transition.
Because of that reason we observe a susceptibility peak at T=Tc.Also we have observed divergence of fluctuation of energy at $T=T_c$ and because of that we observed heat-capacity peak at $T=T_c$.
This peak is increasing with system size and we can observed it (Fig:1) from our calculation very easily.This kind of behaviour easily gives us the idea of the system's collective behaviour in thermodynamic limit.Here we have observed a very nice variation of susceptibility and heat-capacity with T for different system sizes and we are observing this behaviour for two set of lattice size $N=10\times 10 \times 10 $ and $N=5 \times 5 \times 5 $.This clearly indicates that the susceptibility and heat-capacity will diverge at thermodynamic limit.Which is the generic nature of second order phase transition.We know that during phase transition at $T=T_c$ correlation length of the system usually diverge and system follows a collective dynamics and we call this spin dynamics as collective phenomena of the phase transition.
Since the correlation length is diverging so the small local fluctuation can effect the whole system and that is the major reason of appearing peak of heat-capacity and susceptibility at T=Tc.During  phase transition  all spins in a lattice become highly correlated and because of that reason it is very difficult to flip a single spin and the dynamics of the whole system becomes slowed down.We call this phenomena critical slowing down.This critical slowing down situation is one of the major obstacle for observing phase transition using single spin flipping algorithm and because of that more often cluster spin flipping algorithm is used to avoid critical slowing down.However in last few decades because of the high performance computing facilities and using our state art simulation technique we overcome this barrier and we can easily simulate the spin dynamics using single spin flipping technique.Here we have stick to the single spin flipping Metropolis algorithm and that gives us very accurate results.The main advantage of this single spin flipping algorithm is that it is very very simple and easy to simulate any kind of complex lattice system with this kind of algorithm.

\subsection{3D XYZ Spin Model}
We have simulated 3D XYZ spin model in 3D cubic lattice system\cite{kamilov}\cite{jankemonte}.We have been using $5\times 5 \times 5$ and $10 \times 10 \times 10$ cubic lattice grid for the present study.We have been simulated XYZ spin at each lattice site after choosing randomly $\theta$ and $\phi$ coordinate in spherical polar coordinate.We have been  using the following formula for $\theta$ and $\phi$ as $\theta=cos^{-1}(2u_1-1)-\pi$ and $\phi=2 \pi(1-u_2)$.Here $u_1$ and $u_2$ are computer generated random number which can take random value from 0 to 1 uniformly and we have tested its uniformity before the study.From the expression of $\theta$ and $\phi$  we can easily say that if $u_1$ takes value from 0 to 1 randomly then the $\theta$ vary from 0 to $\pi$ and if the random number $u_2$ takes value from 0 to 1 then $\phi$ takes value form 0 to $2\pi$ randomly.In this way we have simulated initial random classical spin configuration to start the simulation process. We have assigned the orthogonal spin component of each spin using the spherical polar coordinate over the unit sphere where the value of spin components $S_x$,$S_y$ and $S_z$ are following $S_x=|S|sin(\theta)cos(\phi)$, $S_y=|S|sin(\theta)sin(\phi)$, $S_z=|S|cos(\theta)$.Here |S| is the magnitude of classical spin moment.So from the  component of classical spin moment $S_x$ ,$S_y$ and $S_z$ we can easily say that  if the $\theta$ and $\phi$

\begin{figure}[htp]
\centering
\includegraphics[scale=0.250]{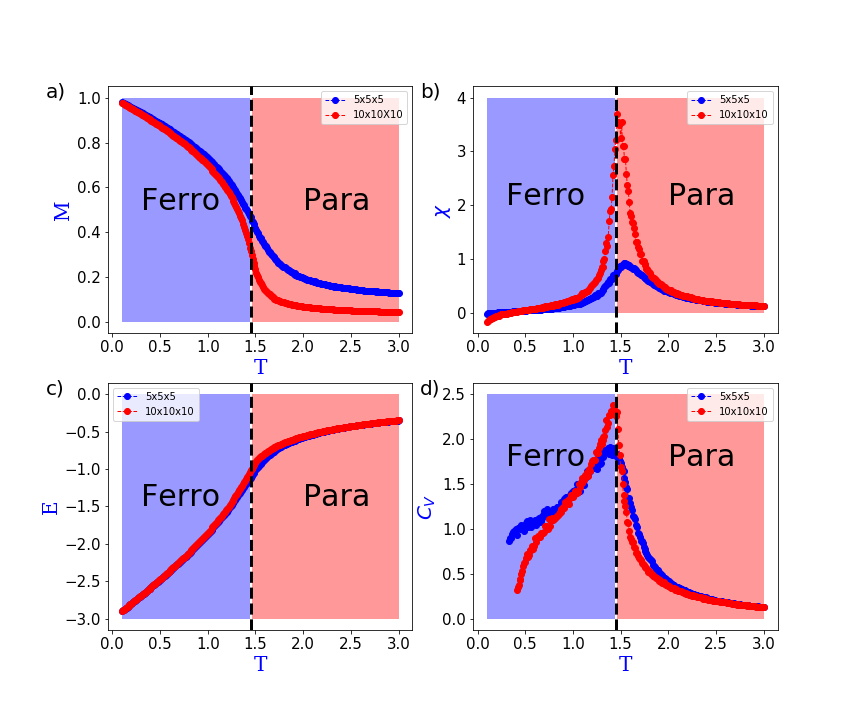}
\caption{a)Magnetization(M) with temperature (T).b)Susceptibility($\chi$) with temperature(T).c)Energy with temperature(T). d)Heat capacity($C_v$) with temperature(T).Here we have presented this results for two different lattice size,which are $5 \times 5 \times 5 $, $10 \times 10 \times 10 $.Here we have presented the results for 3D isotropic XYZ Heisenberg spin model.Here the dotted vertical line indicating phase transition line and here transition temperature($T_c$) is 1.449.}
\label{}
\end{figure}
coordinate  randomly vary with in the prescribed range then that particular classical spin moment will find a unique direction over the unit sphere and each point on the surface of sphere  indicate a distinct state in spin space.Here we have chosen spin magnitude |S| =1 and the value of exchange interaction J=1,which indicate that the ground state of our present system is going to show FM ground state.Here spins are situated on the 3D cubic lattice grid.So each spin has three spatial coordinate and we have assigned our spin component as $S_x=S_x(i,j,k)$,$S_y=S_y(i,j,k)$,$S_z=S_y(i,j,k)$.We have been following here the same Metropolis single spin flipping algorithm as before to equilibrate the system and after reaching the equilibrium we have taken the average of different thermodynamic quantity using $5 \times 10^4$ ensembles,which are just equivalent spin configurations at a particular temperature.We have discarded $5 \times 10^4$ Monte Carlo steps to achieve equilibrium spin configuration state at a particular temperature and after reach that equilibrium state we have taken the average over $5 \times 10^4$ MC steps.Here we have observe from our simulation results that the phase transition takes place at Tc=1.449.We have estimated the Tc from susceptibility and heat-capacity peak.Here we have been noticing finite size scaling behaviour for two set of data $5\times 5 \times 5$ and $10\times 10\times 10$.We have observed that for higher lattice size the peak of the curve is more sharply diverge which basically indicate that those thermodynamic quantity show singular behaviour at T=Tc which is a generic characteristic of second order continuous type of phase transition and scaling behaviour for finite size system.Here we have calculated magnetization(M),where we have  defined magnetization  as 
\begin{eqnarray}
M=\sqrt{m_x^2 +m_y^2+m_z^2}
\end{eqnarray} 
Here $m_x=\sum S_x/L^3$,$m_y=\sum S_y/L^3$ and $m_z=\sum S_z /L^3$.We measure the susceptibility as the fluctuation of magnetization as use the following formula through out our calculation.
\begin{eqnarray}
\chi =L^3(<M^2>-<M>^2)/T
\end{eqnarray}
Here we have been observing the peak at susceptibility curve in our present model system and simulation results and that peak sharply increase with system size.This singular behaviour also clearly indicate a second order phase transition. Also in 3D XYZ model the behaviour of heat-capacity($C_V$)shows singular behaviour and from  our simulation result we have been confirm that.The peak of heat-capacity diverge with system size. Here we show that the peak of heat-capacity of $10\times 10\times 10$ lattice is higher then the peak of $5\times 5\times 5$ lattice.From this heat capacity curve we have eastimated that Tc=1.449.Here we have been observing that the energy of the spin decrease as we cool down system from higher value of T which is above Tc.In this way the energy value reach its lowest minimum value that indicate that system has reach a stable ground state and here we observe ferromagnetic state where all spins are align along a particular direction of unit sphere.This is also expected theoretically because we have set exchange value J=1.From the Fig(2) we can notice the variation of magnetization (M) with T and it's showing a smooth continuous variation.Here M smoothly reach to saturation value which is 1 here.This indicate a generic second order phase transition where order parameter changes smoothly.We have noticed that variation of thermodynamic variables for both $5\times 5\times 5$ and $10\times 10\times 10$ lattice size.We have observed that the variation of $10\times 10\times 10$ is more steeper then $5\times 5 \times 5$,which indicate the scaling behaviour of the system.Here we have been using fluctuation of energy of a spin as measure of heat-capacity.The generic formula we have used for calculating the value of $C_V$ is 

\begin{eqnarray}
C_V=L^3(<E^2>-<E>^2)/T^2
\end{eqnarray} 
Here we have noticed a sharp peak of heat-capacity and that peak increase with system size.The peak is arising around the transition temperature $T_c=1.449$.Here we have considered strictly isotropic model and  exchange interaction between Sx-Sx and Sy-Sy is same that is Jx=Jy=J=1. 
\section{Discussion and Conclusion}
Here we have successfully studied the basic XY and XYZ spin model in simple cubic lattice with classical Monte Carlo simulation.We have noticed the  singular behaviour of  different thermodynamic quantities like susceptibility and heat-capacity near at the transition point.We have observed the spontaneous symmetry breaking in this present model and simulation and study the collective dynamics during the phase transition.We have estimated the transition temperature $T_c$ for 3D isotropic XY and XYZ classical spin model with simple cubic lattice.The value of Tc is very closed to previous reported value and theoretical results.Here we have performed and presented very basic and primary results just to establish our simulation methodology and show that it is successfully able to perform classical Monte Carlo simulation with Heisenberg spin using single spin flipping algorithm.As an example we have studied the phase transition of 3D XY and XYZ Heisenberg spin model and our simulation results very accurately match with the results by cluster spin flipping algorithm.
\section{Acknowledgement:}
N.B is greatly acknowledging Prof.M.Acharyya for introducing the Monte Carlo simulation technique and for numerous important discussion and comments during the progress of this work.Also author is highly thankful to Prof.J.Jung for introducing the Heisenberg spin simulation and for several interesting,deep details and intuitive discussion in this field and several help during the debugging of Monte Carlo code.N.B is thankful to UOS,South Korea for providing initial financial support through SAMSUNG and NRF project.N.B is greatly thankful to IIT Kanpur for giving research position. 
\cite{data}

\end{document}